\documentclass[sigconf]{acmart}
\usepackage{multirow}
\usepackage{tabularx}
\usepackage{amsmath}
\usepackage{amsfonts}
\usepackage{algorithm}
\usepackage{algorithmic}
\usepackage{graphicx}
\usepackage{booktabs}
\usepackage{geometry}
\usepackage{array}
\usepackage{multirow}
\usepackage{float}
\usepackage{cite}
\usepackage{siunitx}
\usepackage{booktabs}
\usepackage{hyperref}
\usepackage{url}
\AtBeginDocument{%
  }

\setcopyright{acmlicensed}
\copyrightyear{2018}
\acmYear{2018}
\acmDOI{XXXXXXX.XXXXXXX}
\acmConference[Conference acronym 'XX]{Make sure to enter the correct
  conference title from your rights confirmation email}{June 03--05,
  2018}{Woodstock, NY}
\acmISBN{978-1-4503-XXXX-X/2018/06}

\begin{document}

\title{AtomGraph: Tackling Atomicity Violation in Smart Contracts using Multimodal GCNs}

\author{Xiaoqi Li}
\affiliation{%
  \institution{Hainan University}
  \city{Haikou}
  \country{China}
}
\email{csxqli@ieee.org}

\author{Zongwei Li}
\authornote{Corresponding author.}
\affiliation{%
  \institution{Hainan University}
  \city{Haikou}
  \country{China}
}
\email{lizw1017@hainanu.edu.cn}

\author{Wenkai Li}
\affiliation{%
  \institution{Hainan University}
  \city{Haikou}
  \country{China}
}
\email{cswkli@hainanu.edu.cn}

\author{Zeng Zhang}
\affiliation{%
  \institution{Hainan University}
  \city{Haikou}
  \country{China}
}
\email{zz1up@hainanu.edu.cn}

\author{Lei Xie}
\affiliation{%
  \institution{Hainan University}
  \city{Haikou}
  \country{China}
}
\email{xielei@hainanu.edu.cn}

\renewcommand{\shortauthors}{Li et al.}

\begin{abstract}
Smart contracts are a core component of blockchain technology and are widely deployed across various scenarios. However, atomicity violations have become a potential security risk.    
Existing analysis tools often lack the precision required to detect these issues effectively. 
To address this challenge, we introduce \textsc{AtomGraph}, an automated framework designed for detecting atomicity violations.
This framework leverages Graph Convolutional Networks (GCN) to identify atomicity violations through multimodal feature learning and fusion. Specifically, driven by a collaborative learning mechanism, the model simultaneously learns from two heterogeneous modalities: extracting structural topological features from the contract's Control Flow Graph (CFG) and uncovering deep semantics from its opcode sequence.
We designed an adaptive weighted fusion mechanism to dynamically adjust the weights of features from each modality to achieve optimal feature fusion.
Finally, GCN detects graph-level atomicity violation on the contract. Comprehensive experimental evaluations demonstrate that \textsc{AtomGraph} achieves 96.88\% accuracy and 96.97\% F1 score, outperforming existing tools. 
Furthermore, compared to the concatenation fusion model, \textsc{AtomGraph} improves the F1 score by 6.4\%, proving its potential in smart contract security detection.
\end{abstract}




\maketitle

\section{Introduction}
In recent years, frequent blockchain security incidents have posed severe challenges to the digital asset ecosystem. Critical infrastructures such as Decentralized Exchanges (DEX), Decentralized Finance (DeFi) protocols, and cross-chain bridges~\citep{gao2025implementation, shen2025blockchain, huangAdvancingWeb302024, aguilarSmartContractFamilies2024} have repeatedly become targets of attacks, causing significant economic losses and trust crises. 
According to statistics from SlowMist Hacked~\citep{slowmist}, the February 2025 theft on the Bybit platform was particularly noteworthy. Over \$1.46 billion was stolen, making it one of the largest cryptocurrency thefts within the past three years.
Behind these massive losses lie the complex and severe security challenges that blockchain systems have long faced~\citep{chen2025chatgpt, gong2025information, zhang2025risk, xiang2025security}. Among the top 10 common attack types, smart contract vulnerabilities dominate, accounting for as much as 26.55\%~\citep{wu2025security, peng2025mining}.
Among these numerous security risks, atomicity violations represent a particularly insidious and destructive smart contract defect. They violate the fundamental principle that transactions must be fully successful or completely fail.
Such defects are difficult to identify through manual audits and also pose significant challenges to traditional automated analysis tools.

To address these challenges, we propose \textsc{AtomGraph}, a novel automated detection framework for atomicity violations in smart contracts.
This framework is based on multimodal GCNs and achieves deep analysis of bytecode through a collaborative learning paradigm for heterogeneous information~\citep{10646885}.
Specifically, \textsc{AtomGraph} first converts the contract into a CFG and learns features from two complementary modalities: 
\underline{(1)} Structural Modality: capturing control flow dependencies from the topological structure of the CFG.
\underline{(2)} Semantic Modality: Understands the execution intent of the code from opcode sequences in basic blocks.
To effectively fuse these heterogeneous features, we innovatively designed an adaptive weighted fusion mechanism that dynamically evaluates and adjusts modal weights, achieving optimal feature combination within a unified embedding space.
Finally, the GCN is used to detect graph-level atomicity violations on contracts.

The main contributions of this paper are as follows:
\begin{itemize}
    \item To the best of our knowledge, we conducted the first systematic study on detecting atomicity violations in smart contracts and propose \textsc{AtomGraph} for automated detection.
    
    \item We designed a novel adaptive weighting fusion mechanism that extracts structural topology and semantic features, dynamically adjusting weights to achieve optimal feature representation.
    
    \item Experiments demonstrate that \textsc{AtomGraph} achieves an F1 score of 96.97\% and outperforms existing tools across multiple metrics. It also exhibits competitive performance in analytical efficiency and practical applications. 
    \item We have released \textsc{AtomGraph}'s codes and experimental data at \url{https://figshare.com/s/9b36cfe6056be2af9146} for reproducibility of our experiments. 
\end{itemize}

\section{Related Work}
\textbf{Smart Contract Vulnerability Detection}.
These methods are primarily categorized into static and dynamic analysis.
Static analysis can be further subdivided into symbolic execution, such as Oyente~\citep{Luu_2016_Making} and Slither~\citep{Feist_2019_Slither}. They also include pattern-matching tools like Securify~\citep{tsankov2018securify} and Ponzi scheme analysis~\citep{zhang2025security}. 
Other methods such as taint analysis with Ethainter~\citep{brent2020ethainter} or intermediate representation analysis using Vandal~\citep{brent2018vandal}, Conkas~\citep{Veloso__Conkas} and MoveScanner~\citep{luo2025movescanner, wang2024ContractsentryStaticAnalysis, wangContractCheckCheckingEthereum2024}.
These methods typically offer advantages in execution speed and high coverage, but they also commonly suffer from high False Positive Rates (FPRs).
Dynamic analysis methods, including Echidna~\citep{grieco2020echidna} and Harvey~\citep{wustholz2020harvey}, detect vulnerabilities by executing code and capturing runtime issues~\citep{li2025penetrating, li2025interaction, liDemoEnhancingSmart2024}. However, they face limitations due to path explosion and difficulty generating test cases.
Additionally, hybrid approaches like Mythril~\citep{mythril} and Manticore~\citep{Mossberg_2019_Manticore} combine static and dynamic analysis. They enhance detection accuracy through mutual feedback, but they generally require more computational resources and time.

Deep learning-based methods demonstrate significant potential~\citep{Zhuang_2020_Smart, ZHU2021107920, peng2025multicfv, wang2025ai, li2025facial}. Tools such as Clear~\citep{chen2024improving} and SCVHunter~\citep{luo2024scvhunter} leverage models to learn vulnerability patterns and Features from large-scale smart contract samples, enabling automated detection and prediction of novel vulnerabilities. 

\textbf{Smart Contract Atomicity Violations}.
The atomicity principle requires that a transaction must possess an all-or-nothing~\citep{zhang2025bounded}. Either all components execute successfully, or all must be rolled back and canceled, preventing any intermediate or partially completed states~\citep{li2024detecting}.
In smart contracts, when a sequence of operations that must be executed as a whole is unexpectedly interrupted or only partially completed, it results in an atomicity violation, leading to inconsistent states~\citep{liu2025detecting, wu2025atomicity, xiao2025parallelizing, caiEnablingCompleteAtomicity2024}.

Common forms of atomicity violations in smart contracts include:
(1) Reentrancy attacks: Malicious re-entry of external calls before state updates complete~\citep{song2025silence}.
(2) Check-Effect-Interaction race conditions: Exploitation of the time window between check conditions and resource usage.
(3) Improper Exception Handling: Partial state modifications occur during exceptions without proper rollback.
(4) Cross-Domain Non-Atomicity: Partial failures across multiple smart contracts or different blockchain systems~\citep{liu2025empirical, huang2025comparative, hanOSwapPreservingAtomicity2026, 10381780}.

\vspace{-0.6em}
\section{Method}
\begin{figure}[h]
\centering
\includegraphics[width=\linewidth]{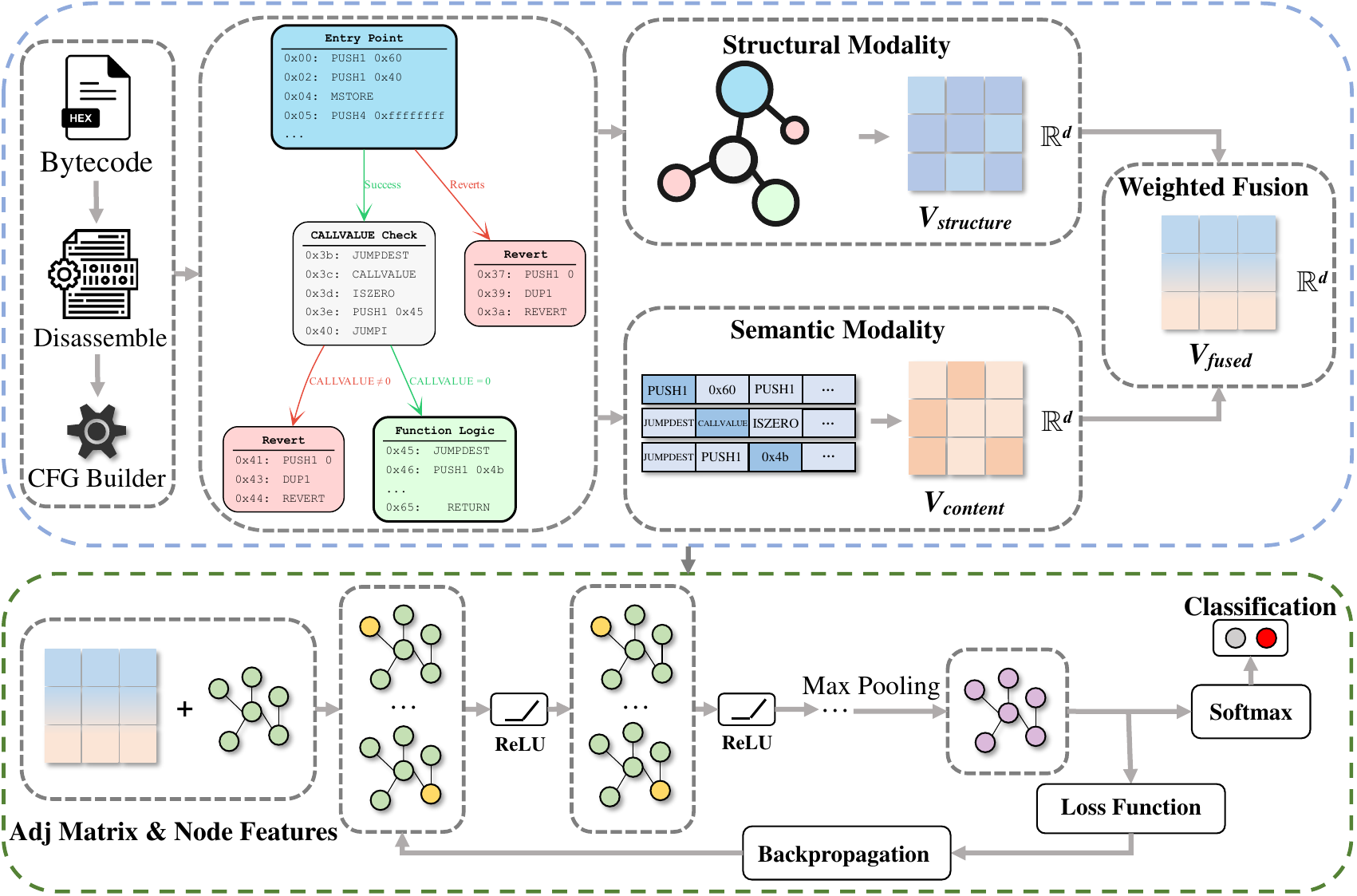}
\caption{Overall architecture of AtomGraph.}
\label{fig:atomgraph_framework} 
\end{figure}

The overview of \textsc{AtomGraph}’s architecture is shown in Figure~\ref{fig:atomgraph_framework}. This architecture consists of three main modules:
\underline{(1)} CFG Construction: Transforms smart contract opcodes into a CFG by defining basic blocks and analyzing jump relationships, outputting a DOT-formatted graph.
\underline{(2)} Adaptive Weighted Fusion: Combine structured and semantic features, and fuse them using optimized weights.
\underline{(3)} GCN Classification: Use GCN to process these fused and normalized features for contract analysis.

\textbf{CFG Construction}.
The CFG is a structured representation of smart contract execution logic, clearly illustrating the basic blocks, branching structures, and control flow transfers within a contract~\citep{zhen2024gnn, arceriSoundConstructionEVM2024}. Building upon existing research, we employ static analysis methods to achieve precise conversion from opcodes to CFGs.
First, we disassembled the raw bytecode stream into an opcode sequence that complies with the Ethereum Virtual Machine (EVM)~\citep{ma2025uncovering}specification. 

Based on disassembled opcodes, we utilize standard basic block partitioning algorithms to identify code structures. Specifically, a basic block is a contiguous sequence of instructions with a single entry point and a unique exit point~\citep{wang2023smart}. Basic block boundaries are defined by specific control flow instructions such as \texttt{JUMP}, \texttt{JUMPI}, \texttt{STOP}, \texttt{REVERT}, \texttt{RETURN}, and the target addresses of jump instructions.
After identifying all basic blocks, we construct the edges of the CFG by analyzing jump relationships between blocks~\citep{10.1145/3699597}. 
For example, when a basic block ends with a conditional jump instruction (\texttt{JUMPI}), two outgoing edges are generated: one pointing to the basic block at the jump target address, and another pointing to the next basic block in sequential execution. 
By systematically connecting all basic blocks, we construct a complete CFG that precisely reflects the underlying execution logic of the contract.

Ultimately, the conversion process generates a graph description file in DOT format. 
This file details the opcode sequences of nodes, the connectivity relationships of edges, and their associated conditions. 

\textbf{Adaptive Weighted Fusion}.
We introduce a multimodal learning paradigm, decoupling smart contract features into semantic and structural modalities. 
The overall fusion procedure is summarized in Algorithm~\ref{alg:atomgraph_weighted_fusion}.
\setlength{\textfloatsep}{0pt}
\begin{algorithm}[ht]
\small
\caption{AtomGraph adaptive weighted fusion algorithm}
\label{alg:atomgraph_weighted_fusion}
\begin{algorithmic}[1]
\REQUIRE Control Flow Graph $G=(V,E)$, weight parameters $\alpha, \beta$
\ENSURE Fused feature representation $\mathbf{V}_{fused} \in \mathbb{R}^{|V| \times d}$

\STATE \textbf{// Step 1: Multimodal Feature Extraction}
\STATE $\mathbf{V}_{structure} \leftarrow$ Node2Vec$(G)$ \COMMENT{Structural feature learning}
\STATE $\mathbf{V}_{content} \leftarrow$ Word2Vec$(G.labels)$ \COMMENT{Semantic feature extraction}

\STATE \textbf{// Step 2: Feature Normalization Preprocessing}
\FOR{$i = 1$ to $|V|$}
    \STATE $\mathbf{v}_{s,i} \leftarrow \frac{\mathbf{V}_{structure}[i]}{\|\mathbf{V}_{structure}[i]\|_2}$
    \STATE $\mathbf{v}_{c,i} \leftarrow \frac{\mathbf{V}_{content}[i]}{\|\mathbf{V}_{content}[i]\|_2}$
\ENDFOR

\STATE \textbf{// Step 3: Adaptive Weighted Fusion}
\FOR{$i = 1$ to $|V|$}
    \STATE $\mathbf{V}_{fused}[i] \leftarrow \alpha \cdot \mathbf{v}_{s,i} + \beta \cdot \mathbf{v}_{c,i}$
    \STATE \textbf{// Constraint}: $\alpha + \beta = 1$
\ENDFOR

\STATE \textbf{// Step 4: Fusion Quality Validation}
\STATE $score \leftarrow$ EvaluateFusionQuality$(\mathbf{V}_{fused})$
\RETURN $\mathbf{V}_{fused}$
\end{algorithmic}
\end{algorithm}

The semantic modality employs Word2Vec to learn opcode sequence semantics, generating semantic embedding vectors $\mathbf{v}_{content}$, while the structural modality employs Node2Vec to learn CFG topology, generating structural embedding vectors $\mathbf{v}_{structure}$.
We employ an adaptive weighted fusion mechanism, performing a linear combination through optimal weight parameters. 

\begin{equation}
\mathbf{v}_{weighted} = \alpha\,\mathbf{v}_{structure} + \beta\,\mathbf{v}_{content},\quad \alpha+\beta=1
\end{equation}

The optimal weights are determined by minimizing the validation set loss function:
\begin{equation}
(\alpha^*, \beta^*) = \arg\min_{\alpha,\beta} \mathcal{L}_{val}(f_{GCN}(f_{fusion}(\mathbf{v}_s, \mathbf{v}_c; \alpha, \beta)))
\end{equation}

To ensure numerical stability, L2 normalization is performed before fusion:
\begin{equation}
\mathbf{v}_{weighted} = \alpha \frac{\mathbf{v}_{structure}}{\|\mathbf{v}_{structure}\|_2} + \beta \frac{\mathbf{v}_{content}}{\|\mathbf{v}_{content}\|_2}
\end{equation}

\textbf{Graph Convolutional Network}.
After obtaining high-quality multimodal node features, we employ GCN as the core classification model, integrating node neighborhood information through multiple GraphConv layers~\citep{shang2025cegt}. The fundamental propagation rule for each layer is as follows:

\begin{equation}
H^{(l+1)} = \sigma\left(\tilde{D}^{-\frac{1}{2}} \tilde{A} \tilde{D}^{-\frac{1}{2}} H^{(l)} W^{(l)}\right)
\end{equation}

Here, $\tilde{A} = A + I_N$ is the adjacency matrix with self-connections added, $\tilde{D}$ is its corresponding degree matrix, $H^{(l)}$ is the feature matrix of the nodes at layer $l$, $W^{(l)}$ is the layer-specific trainable weight matrix, and $\sigma$ is a non-linear activation function.

The AdamW optimizer is employed with a dynamic learning rate scheduling strategy~\citep{10480574}. We integrate focus loss functions, dynamic class weighting, and data augmentation techniques to address class imbalance. Dropout regularization and gradient clipping enhance model generalization and training stability.

\section{Experiments}
All experiments were conducted on a workstation with an NVIDIA GeForce GTX 4070Ti GPU, Intel(R) Core(TM) i9-13900KF CPU, and 128GB RAM. The operating system is Ubuntu 22.04 LTS, and the software environment includes Python 3.9 and PyTorch 2.0.1.



\textbf{Dataset}.
This study's dataset is based on the SmartBugs Wild~\citep{DurieuxEtAl2020ICSE} collection of 47,398 real contracts and DAppSCAN~\citep{10486822, ayubSoundAnalysisMigration2024}, which includes 21,457 contracts and 608 audit reports.
A specialized dataset comprising 1,880 smart contract samples was constructed through systematic screening and manual annotation.
We used 90\% of the dataset for training and 10\% for testing.  
Table~\ref{tab:dataset_distribution} summarizes the smart contract data used.


\begin{table}[h]
\small
\centering
\caption{The collected dataset for our evaluation.}
\label{tab:dataset_distribution}
\begin{tabular}{@{}cccc@{}}
\toprule
\textbf{Classification} & \textbf{Label} & \textbf{Count} & \textbf{Percentage (\%)} \\
\midrule
Normal Contracts           & 0 & 837   & 44.52 \\
Defective Contracts & 1 & 1,043 & 55.48 \\
\midrule[\heavyrulewidth]
\textbf{Total} & \textbf{--} & \textbf{1,880} & \textbf{100.00} \\
\bottomrule
\end{tabular}
\renewcommand{\arraystretch}{1.0}
\end{table}

 
\textbf{Evaluation Metrics.} The effectiveness of \textsc{AtomGraph} is evaluated based on the following research questions (RQs):
\underline{RQ1:} Is \textsc{AtomGraph} capable of accurately identifying atomicity violations in smart contracts on the public dataset?
\underline{RQ2:} Can \textsc{AtomGraph} find atomicity-related violations undetectable by other tools? How does it compare with existing tools?
\underline{RQ3:} How does \textsc{AtomGraph} perform in terms of execution efficiency and system stability compared to existing tools?
\underline{RQ4:} What is the effectiveness of different fusion strategies in \textsc{AtomGraph}'s architecture?
We have established a comprehensive evaluation metric system, including  \textbf{Accuracy (Acc)}, \textbf{Precision}, \textbf{Recall}, \textbf{F1 score}, and \textbf{FPR}.




\begin{table}[h]
\small
\centering
\caption{Performance metrics of AtomGraph.}
\label{tab:atomgraph_performance}
\begin{tabular}{@{}lccccc@{}}
\toprule
\textbf{Tool} & \textbf{Acc(\%)} & \textbf{Recall(\%)} & \textbf{Precision(\%)} & \textbf{F1(\%)} & \textbf{FPR(\%)} \\
\midrule
\textbf{AtomGraph} & 96.88 & 98.27 & 96.00 & 96.97 & 0.05 \\
\bottomrule
\end{tabular}
\end{table}

\textbf{Answer to RQ1: Defects Detection in Dataset.}
Table~\ref{tab:atomgraph_performance} shows the performance of \textsc{AtomGraph} on the atomicity violation detection.
\textsc{AtomGraph} demonstrated remarkable performance in experiments, achieving an accuracy rate of 96.88\% and an F1 score of 96.97\%. 
While maintaining a precision rate of 96.00\%, the model achieves a high recall rate of 98.27\% and strictly controls the FPR at 0.05\%. 
\textsc{AtomGraph} achieves a critical balance between maximizing defect detection coverage and minimizing false positives, representing a key advantage for practical applications.

As its core mechanism, \textsc{AtomGraph} converts bytecode into clearly defined "basic blocks," which represent continuous and branch-free sequences of code execution. Furthermore, \textsc{AtomGraph} integrates the structural dependencies between basic blocks and the operational semantics within each basic block. This enables effective detection of atomicity violations caused by execution flow interruptions or errors in operation sequences, especially during control flow transfers across basic blocks.

\begin{table}[ht]
\small
\centering
\caption{Performance comparison of related tools.}
\label{tab:tool_comparison}
\begin{tabular}{@{}lccccc@{}}
  \toprule
  \textbf{Tool} & \textbf{Acc(\%)} & \textbf{Recall(\%)} & \textbf{Precision(\%)} & \textbf{F1(\%)} & \textbf{FPR(\%)} \\
\midrule
Conkas & 48.97 & 33.80 & 61.90 & 43.73 & 29.50 \\
Ethainter & 44.13 & 0.10 & 12.50 & 0.19 & 0.84 \\
Ethor & 49.82 & 31.33 & 56.76 & 40.38 & 28.28 \\
Honeybadger & 37.16 & 24.04 & 92.80 & 38.19 & 7.82 \\
Mythril & 51.26 & 19.92 & 72.28 & 31.24 & 9.55 \\
Osiris & 34.49 & 18.45 & 96.67 & 30.99 & 2.50 \\
Oyente & 47.36 & 4.40 & 100.00 & 8.43 & 0.00 \\
Pakala & 46.75 & 1.12 & 75.00 & 2.21 & 0.43 \\
Securify & 53.27 & 14.32 & 97.20 & 24.96 & 0.49 \\
Vandal & 51.65 & 44.95 & 58.30 & 50.76 & 40.00 \\
\textbf{AtomGraph} & 96.88 & 98.27 & 96.00 & 96.97 & 0.05 \\
\bottomrule
\end{tabular}
\end{table}

\textbf{Answer to RQ2: Comparative Analysis.}
As shown in Table~\ref{tab:tool_comparison}, we conducted comparative experiments of \textsc{AtomGraph} with other 10 smart contract analysis tools~\citep{he2025VulnerabilityDetectionMethod, sun2025MTVHunterSmartContracts}. 
Symbolic execution tools Mythril, Osiris~\citep{torres2018osiris}, Oyente, and Pakala~\citep{pakala} exhibit the typical characteristics of "high precision, low recall". Oyente achieves 100\% precision but only 4.40\% recall, while Osiris achieves 96.67\% precision but only 18.45\% recall.
This reflects the path explosion problem of symbolic execution when dealing with complex state spaces, leading to insufficient analysis coverage. Pattern-matching tools Securify are similarly constrained by the scope of predefined rules, achieving 97.20\% precision but only 14.32\% recall.
Taint analysis tools Ethainter and intermediate representation analysis tools Vandal show the opposite trend: Vandal relaxes detection conditions to improve recall 44.95\%, but this introduces a high FPR of 40\%.

Additionally, tools like Conkas and Ethor~\citep{schneidewind2020ethor} achieved moderate F1 scores, with Conkas scoring 43.73\% and Ethor scoring 40.38\%. While HoneyBadger~\citep{torres2019art}, which focuses on honeypot detection through symbolic execution expansion, achieves high precision but is limited by its narrow detection scope, resulting in an overall recall rate of only 24.04\%.
\textsc{AtomGraph}'s multimodal architecture overcomes these limitations by integrating structural and semantic modalities, achieving an F1 score of 96.97\% while maintaining a FPR of 0.05\%.

\begin{figure}[h]
\centering
\includegraphics[width=\linewidth]{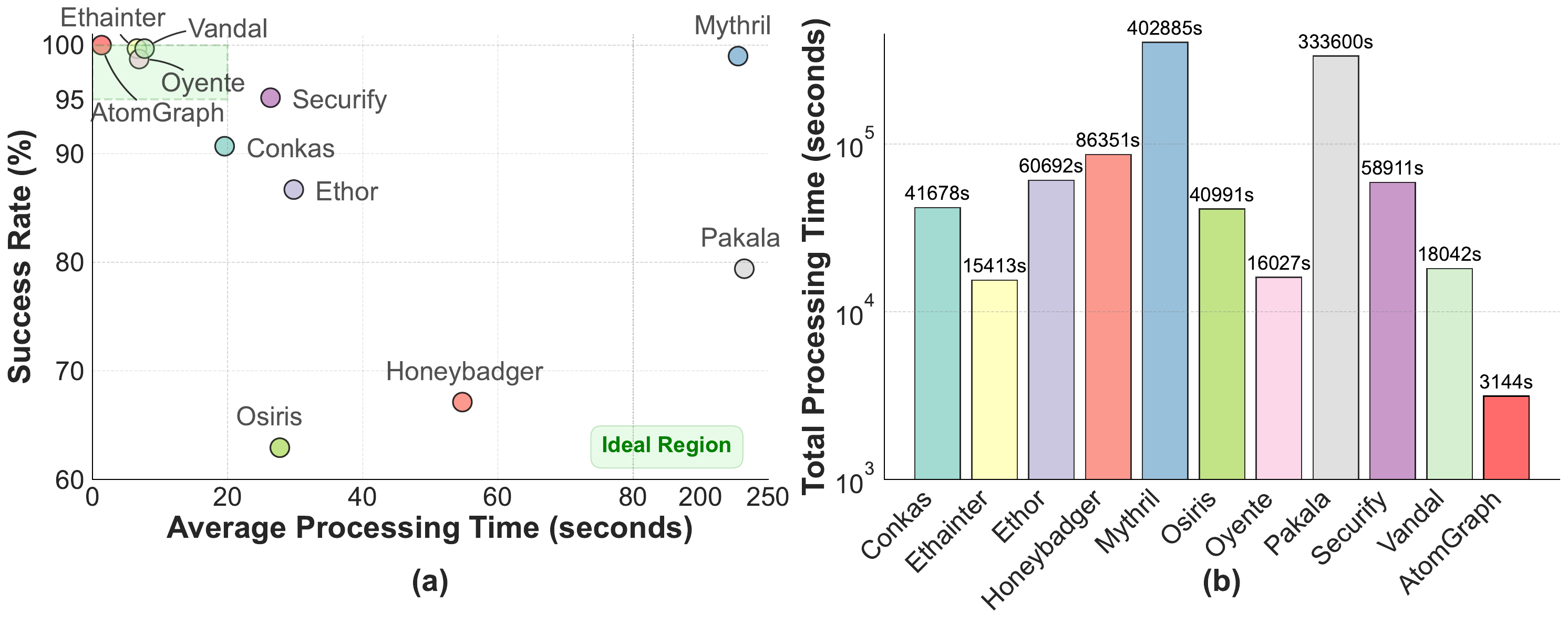}
\caption{Efficiency comparison of related tools. Subfigure (a) illustrates the relationship between average processing time and success rate, while subfigure (b) compares the total processing time across tools.}
\label{fig:efficiency_analysis}
\vspace{-1em}
\end{figure}

\textbf{Answer to RQ3: Efficiency Evaluation.}
We comprehensively evaluated \textsc{AtomGraph}'s execution efficiency and system stability to verify its practicality in large-scale contract analysis. 
Figure~\ref{fig:efficiency_analysis} compares the execution efficiency of different tools.


Experimental results demonstrate that \textsc{AtomGraph} exhibits competitive performance in execution efficiency. Regarding success rate, \textsc{AtomGraph} successfully processed all 1,880 runtime bytecode samples, while other tools achieved success rates ranging from 62.93\% to 99.68\%. 
This disparity primarily stems from differing analytical approaches when handling complex bytecode. 
In processing speed, \textsc{AtomGraph} achieved an average processing time of 1.67 s/contract. This is approximately five times faster than Ethainter's 8.22 s/contract and about 130 times faster than symbolic execution tools like Mythril's 216.49 s/contract and Pakala's 223.44 s/contract.

This efficiency gap stems primarily from \textsc{AtomGraph}'s lightweight feature extraction approach based on pre-trained embeddings, eliminating complex runtime computations. It is important to note that efficiency gains should be evaluated while maintaining detection quality. \textsc{AtomGraph} delivers higher efficiency without compromising detection performance.

\begin{table}[t]
\small
\centering
\caption{AtomGraph ablation study. Baselines: Word2Vec (semantic features) or Node2Vec (structural features) only. Graph augmentation (Graph Aug.): node label content-to-structure conversion and virtual edges via opcode similarity. Fusion strategies: direct concatenation or average weighting.}
\label{tab:ablation_study}
\begin{tabular}{@{}lccccc@{}}
  \toprule
  \textbf{Tool} & \textbf{Acc(\%)} & \textbf{Recall(\%)} & \textbf{Precision(\%)} & \textbf{F1(\%)} & \textbf{FPR(\%)} \\
\midrule
Word2vec  & 82.08 & 85.12 & 85.60 & 85.19 & 0.23 \\
Node2vec  & 89.79 & 95.65 & 88.20 & 91.56 & 0.17 \\
Graph Aug. & 87.92 & 95.22 & 83.99 & 89.15 & 0.20 \\
Concat. Fusion & 92.60 & 95.53 & 86.96 & 90.57 & 0.11 \\
Avg. Fusion & 94.79 & 98.15 & 93.75 & 95.79 & 0.10 \\
\textbf{AtomGraph}   & 96.88 & 98.27 & 96.00 & 96.97 & 0.05 \\
\bottomrule
\end{tabular}
\end{table}
\textbf{Answer to RQ4: Ablation Study.}
We conducted an ablation study using consistent metrics and hyperparameters to assess simplified variants. 
As shown in Table~\ref{tab:ablation_study}, ablation experiments validate the effectiveness of multimodal feature fusion. Among single-feature methods, Word2Vec semantic features achieved an F1 score of 85.19\%, while Node2Vec structural features performed better at 91.56\%. In the comparison of feature fusion strategies, graph augmentation 89.15\%, concatenation fusion 90.57\%, and average fusion 95.79\% all outperformed the single-feature methods. Among these, average fusion demonstrated the best performance but still has room for improvement.

\textsc{AtomGraph}'s adaptive weighted fusion mechanism further enhances performance beyond average fusion, boosting the F1 score from 95.79\% to 96.97\% while reducing the FPR to 0.05\%. This improvement stems from the strategy's ability to assign appropriate weights based on feature contributions, avoiding information dilution inherent in simple averaging. 
Experiments demonstrate that reasonable weight allocation in atomic violation detection tasks helps fully leverage multimodal features' complementary nature.
\section{Future Plans}
Future research will focus on deepening \textsc{AtomGraph} from both theoretical and practical perspectives. Theoretically, we will formally analyze the relationship between CFG topology and atomicity violation patterns, and mathematically characterize optimal fusion weights across different contract types and complexity levels. 
Practically, we will conduct vulnerability mining on real-world deployed contracts, validating its efficacy and reliability by successfully discovering new vulnerabilities and applying for CVE identifiers. Additionally, we will conduct an in-depth analysis of false positive and false negative cases in detection, dissecting their root causes to provide critical insights for model iteration and optimization.

\section{Conclusion}
We propose \textsc{AtomGraph}, a multimodal GCN framework for detecting atomic violations in smart contracts. It constructs precise CFGs from runtime bytecode and extracts features from semantic and structural modalities. An adaptive weighted fusion mechanism maps these into a shared embedding space, followed by GCN-based graph-level classification.
Comprehensive experiments show that \textsc{AtomGraph} achieves outstanding performance, with an accuracy rate of 96.88\% and an F1 score of 96.97\%, outperforming existing tools. Compared to single-modal Word2Vec and concatenation-based baselines, \textsc{AtomGraph} achieves F1 score improvements of 5.41\% and 6.4\%, respectively. It analyzes contracts in 1.67 seconds on average, balancing high recall, low false positives, and high throughput. Overall, \textsc{AtomGraph} provides a potential solution for detecting atomicity violations in smart contracts.

\bibliographystyle{ACM-Reference-Format}
\bibliography{ref}










\end{document}